\documentclass[twocolumn,prl,showpacs,preprintnumbers,amsmath,amssymb,superscriptaddress]{revtex4}

\usepackage{graphicx}
\usepackage{dcolumn}
\usepackage{bm}

\begin{document}


\title{The influence of the long-lived quantum Hall potential on the characteristics of quantum devices\\}

\author{M. Pioro-Ladri\`{e}re}
\affiliation{Institute for Microstructural Sciences, National Research Council of Canada, Ottawa, Ontario, Canada K1A~0R6}
\affiliation{Centre de Recherche sur les Propri\'{e}t\'{e}s \'{E}lectroniques de Mat\'{e}riaux Avanc\'{e}s, Universit\'{e} de Sherbrooke, Sherbrooke, Qu\'{e}bec, Canada J1K~2R1}

\author{A. Usher}
\affiliation{Quantum Interacting Systems Group, School of Physics, University of Exeter, Stocker Road, Exeter, EX4 4QL, U.K.}

\author{A. S. Sachrajda}
\author{J. Lapointe}
\author{J. Gupta}
\author{Z. Wasilewski}
\author{S. Studenikin}
\affiliation{Institute for Microstructural Sciences, National Research Council of Canada, Ottawa, Ontario, Canada K1A~0R6}

\author{M. Elliott}
\affiliation{School of Physics and Astronomy, Cardiff University, 5 The Parade, Cardiff, CF24 3YB,  U.K.}

\date{\today}

\begin{abstract}
Novel hysteretic effects are reported in magneto-transport experiments on lateral quantum devices. The effects are characterized by two vastly different relaxation times (minutes and days). It is shown that the observed phenomena are related to long-lived eddy currents. This is confirmed by torsion-balance magnetometry measurements of the same 2-dimensional electron gas (2DEG) material. These observations show that the induced quantum Hall potential at the edges of the 2DEG reservoirs influences transport through the devices, and have important consequences for the magneto-transport of all lateral quantum devices.\end{abstract}

\pacs{73.23.-b, 73.23.Hk, 73.23.Ra, 73.43.-f}

\maketitle
Electrostatically defined nanostructures are ideal for applications in which tunability is a requirement. Recent advances in quantum-dot designs have resulted in lateral quantum dots which can be tuned down to a single confined electron \cite{ciorga} making such devices candidates for the implementation of either spin or charge qubits, required for quantum computing. The leads for these systems are formed from the 2-dimensional electron gas (2DEG) exterior to the device. The two-dimensional nature of these electron reservoirs is known to have consequences for quantum dot investigations \cite{ciorga,andy}. However, one of the most intriguing properties of 2DEGs, the quasi-dissipationless nature of the quantum Hall effect (QHE), has yet to be considered in relation to these devices. This is the topic of this letter. Sweeping the magnetic field close to integer Landau-level filling factors ($\nu$) induces long-lived eddy currents. Such currents were first detected in 2DEGs by torsion-balance magnetometry \cite{eisenstein} and have been employed as a contact-free technique for the study of the high-current breakdown of both the integer \cite{jones,matthews} and fractional \cite{watts,matthews2} QHEs. Eddy currents have also been detected recently using metallic single-electron transistors as very sensitive electrometers above the 2DEG \cite{huels,klaffs}. These latter experiments provide evidence that the currents exist in devices with ohmic contacts and should therefore be present in the leads of lateral nanostructures. The induced currents can persist for hours, even days after a magnetic field sweep is stopped \cite{jones2}. In this paper we reveal hysteretic effects in the properties of lateral quantum dots and quantum point contacts (QPC). To confirm that these effects are related to long-lived eddy currents, we have also performed torsion-balance magnetometry experiments on the 2DEG using a bulk piece of the same wafer. The induced currents generated within the 2DEG contacts in the QHE regime result in long-lived Hall voltages which influence the electrostatics of both the QPC and the quantum dots. The Hall voltage, and hence the observed effect in the tunnelling spectrum, changes sign when the direction of the field sweep is reversed leading to the hysteretic effects.

The devices were fabricated from a 2DEG within a standard GaAs/(Al,Ga)As single heterojunction. As grown, the 2DEG material had an electron density of $1.7 \times 10^{15}\,\mathrm{m}^{-2}$ and an electron mobility of $200\,\mathrm{m}^2 \,\mathrm{V}^{-1} \,\mathrm{s}^{-1}$ at 4\,K. Measurements were made both on an isolated QPC (inset to Fig. 1(a)) and on a quantum dot incorporating a QPC charge sensor. The circular quantum dot structure adjacent to the QPC in the first device plays no role in the measurements presented in this paper. A QPC in close proximity to the dot of the second device was used as a charge sensor using the approach first reported by Field \textit{et al}. \cite{field}. The detector was biased to a point in its characteristic where the conductance was very sensitive to the local charge environment. Such a charge detector can detect changes in the number of electrons in the quantum dot as a gate voltage is swept and hence measure the Coulomb-blockade addition spectrum. Transport through the QPCs was measured using standard low-frequency AC techniques.

Figures 1 and 2 show the results from the isolated QPC device. In a typical charge detection scenario QPCs are set so that the conductance lies below the last conductance plateau, i.e. below $2e^2/h$. Figure l(a) shows a typical sweep of magnetic field ($B$) taken under such conditions. Both up and down sweeps are shown. In addition to the expected $1/B$ oscillations at low $B$, associated with the Shubnikov-de-Haas effect in the 2DEG contacts, a clear hysteretic effect can be seen around 3.6\,T which corresponds to $\nu=2$ in the 2DEG. A similar effect was also observed at $\nu=1$ but at no other filling factors. Figure 1(b) shows close-ups of typical results at $\nu=2$ and $\nu=1$. Figure 2 shows three further characteristics of the hysteretic features. Figure 2(a) reveals the effect of varying the sweep rate. The size of the features begins to saturate for sweep rates greater than $0.1\,\mathrm{mT}\,\mathrm{s}^{-1}$. Figure 2(b) illustrates that the relaxation time on stopping the sweep has two components, an initial fast decay, with a time constant of about one minute, followed by a much slower decay. The inset shows the dependence of the time constant for the fast decay on filling factor around $\nu=2$; the time constant peaks at exact integer $\nu$, and falls to zero when the filling factor deviates by $\pm0.2$. Finally, the temperature dependence of the effect is shown in figure 2(c). As the temperature is increased the hysteresis is replaced by a step feature related to the chemical potential variation in 2DEGs \cite{ciorga}.

\begin{figure}
\includegraphics{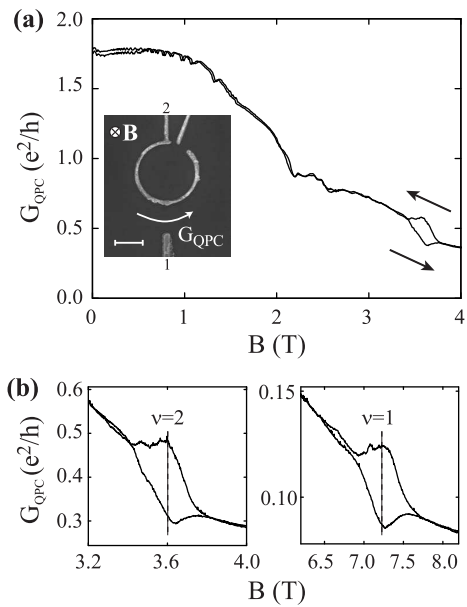}
\caption{Quantum point contact measurements: (a) QPC conductance $G_{QPC}$ as a function of magnetic field $B$, showing the hysteretic feature at 3.6\,T, corresponding to Landau-level filling factor $\nu=2$ in the 2DEG leads. The sweep rate is $0.17\,\mathrm{mT}\,\mathrm{s}^{-1}$, the temperature is 30\,mK. Inset: electron micrograph of the device. The scale bar is 300\,nm. Gates 1 and 2 form the QPC - for these experiments the quantum dot enclosed inside gate 2 plays no role; (b) Enlargement of the hysteretic feature at $\nu=2$ (left), and the same feature at $\nu=1$ (right).}
\label{figure 1}
\end{figure}

\begin{figure}
\includegraphics{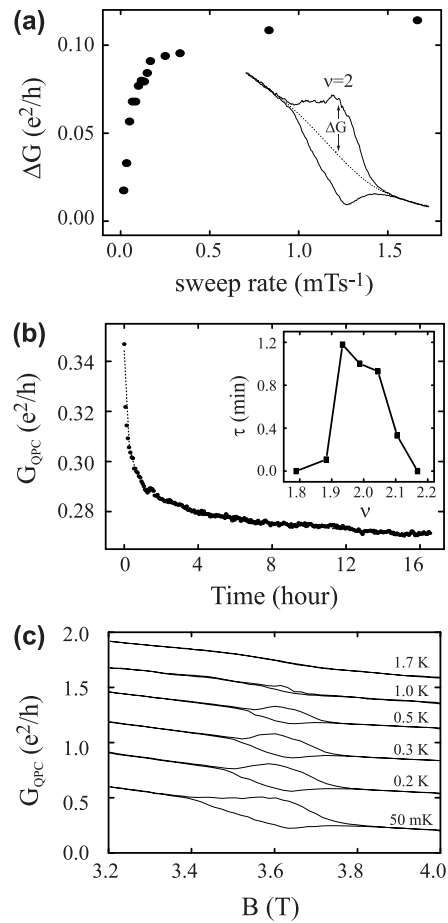}
\caption{Characteristics of the QPC hysteretic feature : (a) Sweep-rate dependence of the hysteretic feature at $\nu=2$, showing a saturation in the peak height $\Delta G$ at sweep rates above about $0.1\,\mathrm{mT}\,\mathrm{s}^{-1}$; (b) Decay of the feature with time when the sweep is stopped at 3.7\,T, showing an initial fast decay followed by a much slower one. The inset shows the dependence of the time constant of the fast decay on filling factor; (c) The temperature dependence of the $\nu=2$ hysteretic feature (successive traces are shifted for clarity).}
\label{figure 2}
\end{figure}

We have also performed torsion-balance magnetometry experiments, following the method described in Matthews \textit{et al}. \cite{matthews3}. An unprocessed 10\,mm by 5\,mm piece of the 2DEG wafer was placed on the rotor of the magnetometer, with the normal to the 2DEG subtending an angle of $20^{\circ}$ with respect to an applied magnetic field of up to 19\,T. The magnetometer was cooled in a low-vibration dilution refrigerator. Figure 3(a) shows the magnetometer output as the magnetic field was swept from 0 to 10\,T, and back, at a rate of $3\,\mathrm{mT}\,\mathrm{s}^{-1}$. Large hysteretic eddy current peaks are observed at $\nu=1$ and 2, in agreement with the QPC experiments while much smaller hysteretic features are also observed at $\nu=3$, 4 and 6. The twin peak structures observed are common in magnetometry measurements on large-area samples, and are believed to be inhomogeneity related. Some less pronounced structure is also observed in the QPC measurements. Figure 3(b) shows the decay of the $\nu=2$ eddy current measured using the torsion-balance magnetometer. After determining the background level, the decay was initiated by sweeping the magnet to the magnetic field corresponding to $\nu=2$, from 0.15\,T above, at a rate of $1.6\,\mathrm{mT}\,\mathrm{s}^{-1}$. The magnitude of the eddy current was then measured as a function of time. For the first 50 seconds, the decay is approximately exponential, with a time constant of 180\,s. From about 100 seconds onwards a much slower power-law decay sets in. The details of this decay behaviour will be the subject of further investigations, but the two distinct timescales associated with the decay mirror our observations in the QPC experiments (Fig. 2(b)). The filling factor dependence of the time constant around $\nu=2$ associated with the initial fast decay is shown in the inset to Fig. 3(b), and again resembles closely the behaviour of seen in the QPC (Fig. 2(b) inset). Figure 3(c) shows the temperature dependence of the eddy currents in the magnetometry experiments, from 39\,mK to 750\,mK. Comparison with Fig. 2(c) once again shows good agreement. The final point of comparison between the magnetometry and the QPC experiments is their behaviour as a function of sweep rate. Numerous previous magnetometry studies \cite{jones,matthews,watts,matthews2} have shown that there is a saturation in the eddy current as the sweep rate is increased, as observed in the QPC measurements. This is caused by an increase in dissipation at high currents accompanying the breakdown of the QHE. This increase in dissipation is indeed also responsible for the fast initial eddy current decay mentioned above. 

\begin{figure}
\includegraphics[scale=1.0]{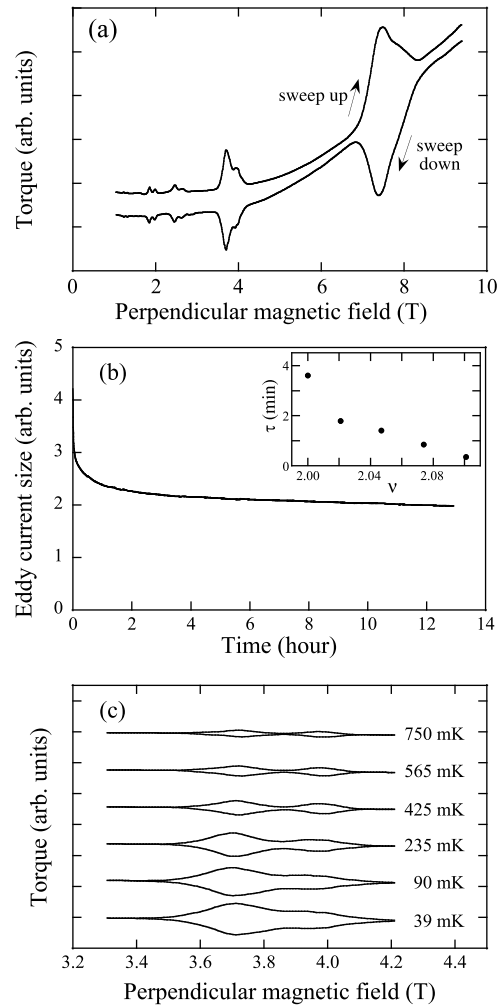}
\caption{Magnetometry measurements: (a) torque (proportional to induced current) as a function of magnetic field, sweep rate, $3\,\mathrm{mT}\,\mathrm{s}^{-1}$, temperature, 45\,mK; (b) Decay of the induced eddy current at $\nu=2$ and (inset) the behavior of the fast-decay time constant around $\nu=2$; (c) Temperature dependence of the $\nu=2$ eddy current (successive traces are shifted for clarity).}
\label{figure 3}
\end{figure}

A very different long-lived hysteretic effect has previously been observed as a function of bias in QPC experiments \cite{wald} at $\nu=2$, related to inter-edge-state spin relaxation and the Overhauser effect. In our experiments we observe the features at low bias not only at $\nu=2$ but also at $\nu=1$ ruling out the Overhauser effect as an explanation of our observations. Given the similarities between the QPC and torsion-balance magnetometry results we conclude that eddy currents are responsible for the hysteretic effects observed in the QPC experiments. Our explanation of the observations is as follows:  eddy currents are induced in both of the 2DEG leads of the QPC. This results in a Hall voltage between the centers of each of these regions and their respective edges. The QPC is in close proximity to the edges of both of these regions, and so the electrochemical potential of electrons passing through the QPC will be altered by an amount related to this Hall potential. Since the Hall potential reverses sign when the sweep direction is reversed, this results in the hysteretic features observed. It is also interesting to note that the detection of eddy currents by a QPC confirms that the eddy currents flow near the edges of the sample, as predicted by Klaffs \textit{et al}. \cite{klaffs}.

Figure 4 demonstrates that the induced Hall potential influences the properties of other lateral nanostructures. A QPC charge detector set to a conductance of less than $2e^2/h$ is used to measure the charge of a quantum dot in the device in the inset of Fig. 4(a). Only one of the dots is activated for these measurements. Figure 4(a) shows the derivative of the current through the charge detector with respect to the voltage on one of the confining gates of the dot (transconductance). The magnetic field was stepped between 4 and 0\,T, and at each step, the gate voltage was swept from 0 to $-0.3\,V$. Thus the occurrence of hysteretic features in these data relies upon the long lifetime of the currents compared to the timescale of the measurements. The dark lines represent the Coulomb-blockade-related features as the fifth to the ninth electrons are added to the dot. The number of electrons is indicated by $N$ in this figure. For these low electron numbers there are very few features associated with level crossings of the quantum dot, and the upward steps in Fig. 4(a) are the well-understood 2DEG chemical potential oscillations in the leads \cite{ciorga} associated with magnetic depopulation of Landau levels. These are labelled with 2DEG Landau level filling factors  $\nu=$ 18 to 4. Since these are 2DEG related features they occur at the same magnetic fields for all the dark lines. At $\nu=2$, however, strikingly different behaviour is observed: a distinct dip occurs along the Coulomb blockade lines. Figure 4(b) reveals that this anomalous feature has the same hysteretic behaviour as in the QPC characteristics of Fig. 1. The results of up and down sweeps are superimposed. Figure 4(c) shows that the same hysteretic behaviour is also present at $\nu=1$. Figure 4(d) confirms that the QPC detector itself is independently influenced by the Hall potential even when the field is being stepped. The figure shows the transconductance as a function of $B$, away from any Coulomb blockade features (the data are extracted from the field stepped curves that make up Fig. 4(b)). Exactly the same hysteretic behavior is seen as in Fig. 1(b). 

\begin{figure}
\includegraphics{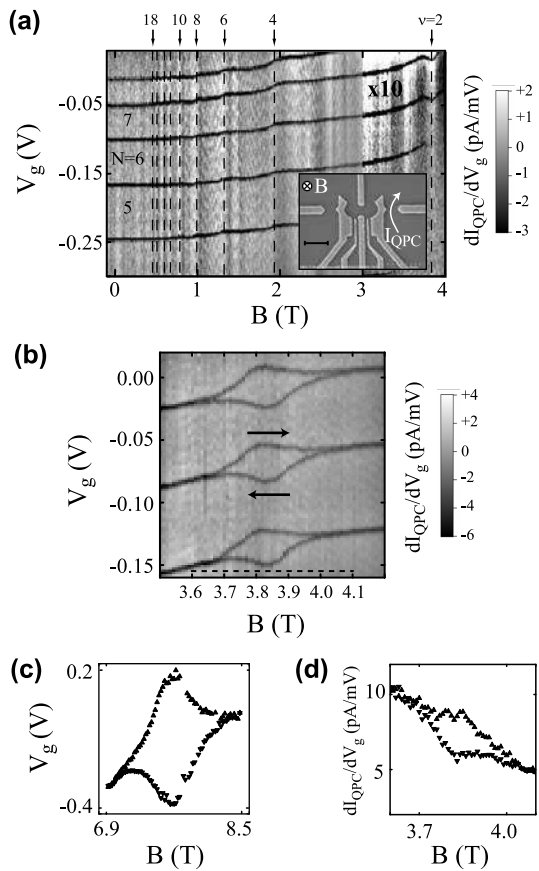}
\caption{Quantum dot measurements: (a) Part of the addition spectrum of the quantum dot obtained by adding the 5th to 9th electron and measured using a QPC charge detector. The dark lines map out the voltage at which an electron is added to the dot. An anomalous dip can be seen at filling factor $\nu=2$. Inset: electron micrograph of the device: only one quantum dot is activated for these measurements. The scale bar is 500\,nm; (b) Detail of the hysteretic behavior; (c) Hysteretic behavior occurring at $\nu=1$; (d) QPC transconductance vs. magnetic field at a gate bias of $V_{g}=-0.16\,V$, indicated by the dashed line in (b).}
\label{figure 4}
\end{figure}

Our interpretation of the hysteretic Coulomb blockade feature is that the Hall potential accompanying the induced eddy currents in the 2DEG leads raises or lowers the local electrochemical potential in the leads, so that the gate voltage at which the electrochemical potentials in the dot and leads are matched becomes dependent on the direction of field stepping. These data confirm that in effect the eddy currents are acting as contactless gates on the quantum dot.

We have observed striking hysteretic features in QPCs and few-electron lateral quantum dots. The role of long-lived induced currents as the origin of these effects was confirmed by comparison with the eddy current properties observed in torsion-balance magnetometry studies of the same 2DEG material. These observations have profound implications both for measurements of the QHE and of lateral nanostructures in magnetic fields. In QHE experiments, it is clear that these induced currents are present, and persist for days, even when the magnetic field is stepped, swept extremely slowly or is static. Furthermore, for sweep rates above $\approx0.1\,\mathrm{mT}\,\mathrm{s}^{-1}$ the induced currents are large enough to cause breakdown of the QHE. Our technique, using nanostructures to probe a 2DEG, provides new opportunities for the study of these effects. It is now evident as a result of this work that induced currents also play a role in experiments on nanostructures, and must be taken into consideration when interpreting measurements in quantizing magnetic fields of devices involving QPCs or quantum dots, such as interferometers. If the effects of these long-lived currents are to be avoided, experiments need to be warmed up to 2\,K after the field sweep is stopped. Finally, our results provide exciting possibilities for contactless control of such lateral nanostructures, for instance the addition of single electrons to quantum dots.

We wish to acknowledge discussions with D.G. Austing and the assistance of P. Zawadzki. MPL acknowledges assistance from NSERC and FQRNT. A.S.S acknowledges support from CIAR. A.S.S and M.E. acknowledge support from CERION programme.


\end{document}